# Experimenting with Experimentation: Rethinking The Role of Experimentation in Educational Design

MOHI REZA, University of Toronto, Canada
AKMAR CHOWDHURY, University of Toronto, Canada
AIDAN LI, University of Toronto, Canada
MAHATHI GANDHAMANENI, University of Toronto, Canada
JOSEPH JAY WILLIAMS, University of Toronto, Canada

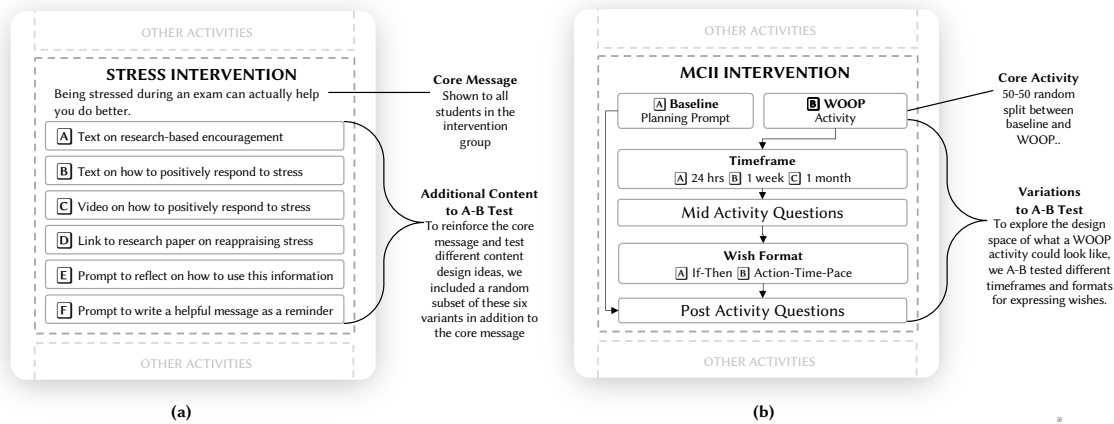

Fig. 1. Summary of Intervention Design for the Stress Reappraisal and MCII Activities

What if we take a broader view of what it means to run an education experiment? In this paper, we explore opportunities that arise when we think beyond the commonly-held notion that the purpose of an experiment is to either accept or reject a pre-defined hypothesis, and instead, reconsider experimentation as a means to explore the complex design space of creating and improving instructional content. This is an approach we call *experiment-inspired design*. Then, to operationalize these ideas in a real-world experimentation venue, we investigate the implications of running a sequence of interventions teaching first-year students "meta-skills": transferable skills applicable to multiple areas of their lives, such as planning, and managing stress. Finally, using two examples as case studies for meta-skills interventions (stress-reappraisal and mental contrasting with implementation intentions), we reflect on our experiences with experiment-inspired design and share six preliminary lessons on how to use experimentation for design.

## 1 INTRODUCTION

Often, randomized experimentation is used in education research to evaluate an intervention, or test a specific hypothesis about teaching or learning. What if we take a creative perspective on the *goals* and *uses* of experiments than what is typically presented, and instead reconsider what value can be gained by thinking in terms of experiments, for the purposes of instructional design, if the principal goal is *not* to analyze the experiment?

Drawing analogy from design, this creative perspective is akin to *appropriating* [7] concepts from the process of designing an experiment and using it for other purposes that are not part of the original intent. [7, 8]. While this may seem unusual, or perhaps even radical, such an effort is in alignment with a fundamental, and arguably human approach to design [7], an "abductive leap from a simple observation about some object to a different vision for how that object might be used to achieve a goal."[7, 8]

In this paper, we try to (re) characterize some of the **constraints** and **affordances** of the experimental approach, and explore the **goals** and **opportunities** at hand for those who engage in or can benefit from experimentation. We draw from our extensive experiences running several A-B interventions over the past three years with thousands of students in real-world university classrooms, and introduce some practical considerations and guidelines that we have found useful when treating such interventions as a novel *experiment-inspired-design process*.

## 2 META-SKILLS INTERVENTIONS AS A VENUE FOR DOING EXPERIMENT-INSPIRED DESIGN

To explore an experiment-inspired design process, we chose an introductory programming course at a large and diverse North American university as our venue for experimentation. This course is offered regularly and has high enrolment rates, with over 1000 students.

Over the past three years, we investigated the effects of a series of randomized interventions intended to teach first-year computer science students "meta-skills": generalizable skills that can help them succeed in multiple areas of life. Examples of meta-skills include learning effective planning, setting goals, managing stress, etc.

We chose these meta-skills topics as our area of investigation because they have some correspondence to past experiments that have shown that they can be effective, meaning that we can shift our focus away from testing these topics' efficacy, and instead, focus on more divergent design goals, such as figuring out how to communicate these ideas to students, and how these interventions fit into the fabric of their multi-faceted lives.

For the purposes of this paper, we will use the following two meta-skills interventions as case studies for the experiment-inspired design process:

(1) **Stress Reappraisal**: the belief that being stressed during an exam can actually help you do better, a concept explored in a class of prior studies [1, 2, 9, 12] in different contexts.

(2) **Mental-Contrasting with Implementation Intentions (MCII)**:, a strategy for increasing follow-through on goals by outlining and visualizing the outcomes and obstatcles before formulating the plan. [5, 13]

Figures 1 (a) and (b) gives a high-level overview of some of the components of these two interventions. We will refer to these interventions when discussing various facets of our approach to experimentation, and will refrain from going into too much detail on the specifics of the results of these interventions as that is beyond the scope and focus of this paper. In the next section, we discuss *experiment-inspired design* by contrasting it with *traditional experiments*. We use the "traditional experiments" terminology to encapsulate existing practices surrounding experimentation where the goal is often restricted to either rejecting or validating some pre-determined and well-defined hypothesis.

## 3 SIX LESSONS FROM REFLECTING ON EXPERIMENT-INSPIRED DESIGN

What if the primary goal of an experiment is *not* restricted to doing a statistical test for checking whether the results are significant? What if our goal is not just to test a theory, but rather, to explore the design space of how to apply that theory in the real world? Deeply consider the divergent *affordances* of experiments when the primary goal is to design and improve instructional content, and draw practical considerations from our experience with designing, deploying, and analyzing meta-skills interventions over several years. We boil these reflections down to 6 preliminary lessons in experiment-inspired design, which largely falls under two major themes: **iterative design** and **divergent thinking**.

## 3.1 Iterative Design & Parallel Prototyping

In this first theme, we reflect on the role of experimentation as a tool for parallel prototyping and continual improvement. We abandon the notion of a singular experiment and instead, think of the opportunities that arise when we focus on a sequence of successive experiments where we try to improve and generate instructional content over time.

(1) **Generate multiple alternative designs through parallel prototyping** [4]. In the Stress Reappraisal intervention case study, we introduced a core stress reappraisal message to students, and A-B tested additional content (see figure 1) reinforcing the core message using six mini factorials [3]. We played with different mediums for reinforcing the core message, e.g. a video where a professor smiles and speaks to his students, a link to a research article with evidence that reappraisal works, prompts to reflect on how to use the idea moving forward, and so on. We found that our intervention had a significant positive effect on exam scores ($p = 0.003$, $d = 0.252$) regardless of gender, and had a larger impact on first-year students, but checking for this significant result was not the primary goal of running the intervention. Instead, our goal was to explore the design space of how to apply stress reappraisal theory in the real world by parallel prototyping [4], and producing multiple ways to communicate a powerful idea to students.

(2) **Iterate on alternative designs over time** [6]. We see our interventions as vehicles for instructional content that can be improved over time. Because our experimental infrastructure is set up for flexibility, we can re-deploy these interventions multiple times – to the same group of students at different points in the term, or to multiple groups of students in future offerings of the same course – and in doing so, add new variations of content as arms in our A-B comparisons as we think of new ideas. For instance, in our MCII case study, our first version of the intervention closely matched a standard mental contrasting activity Oettingen refers to as *WOOP* [5, 10]. It is an acronym for the steps that students use to identify and fulfill their wishes: identifying your **W**ish, imagining the **O**utcome, anticipating the **O**bstacle, and developing a specific **P**lan. [11] After this first version, we could utilize the examples of implementation intentions generated by students from the first version to help students in future versions of the intervention.

(3) **Always ask for user feedback.** At the end of every intervention, we have found it helpful to always ask for student feedback on what worked well or could be improved from the user perspective. We ask users what they think of the activity, how it might or might not be useful to them, and how it can be improved. We also ask them to rate the helpfulness of the activity using a 7-point Likert scale (strongly agree to disagree). We want to go beyond testing whether our interactions impact behaviour; we want to practice iterative design. We do not think of our interventions as single experiments; instead, we set up successive experiments to help us continually improve instructional content. Understanding how users engage with multiple versions of instructional content by soliciting qualitative data on the perception and interpretations of our interventions has helped us produce new ideas to prototype in divergent ways, and improve existing prototypes as we converge to better designs.

## 3.2 Divergent Thinking

In this second theme, we rethink the role of constraints (elements or rules) associated with experimentation, even when those constraints may seem obvious or even sacrosanct. For instance, when doing a between-subjects A-B comparison, one might think it bizarre to give access to *both* versions to the user. One might also think that doing an experiment is worthwhile only if the analysis of the data yields significant effects. Using these examples, we reflect on the benefits of widening or breaking constraints and imposing or appropriating constraints that can push us to engage in more divergent thinking.

(4) **Widening Constraints to factor in diverse stakeholder needs:** Consider the first idea (the A-B comparison). After several discussions with course instructors when creating our meta-skills interventions, we realized that instructors dislike randomizing different versions of resources to different students because questions about fairness arise – what if, by doing such experiments, some students cannot benefit from the "best" version because they were unintentionally assigned a sub-optimal version? To resolve this issue, we widened constraints by giving the *option* to access *all versions* of our intervention at the end of each meta-skill module, being *transparent* about there being multiple versions, and giving students the autonomy to choose whether they want to see all versions. We kept track of who chose to access all versions so that we could then factor this information into our analysis. In this case, we took into account the need of the instructor to be fair, the need of the student to have access to the best resources, and the need of the researcher to do meaningful experimentation.

(5) **Widening Constraints to Normalize Experimentation:** Another reason to widen a constraint may be to think about the benefits of running an imperfect or incomplete experiment. Revisiting our second constraint (what is a worthwhile experiment), it may seem obvious. However, when doing sequences of experiments at scale conducted over several years, we have realized that running experiments can be worthwhile *even if you may not have the resources available to analyze the results* immediately after deployment. For instance, we found that the act of normalizing regular experimentation in a course has long-term positive effects, such as when trying to systematically think about the impact of the resources we create and to keep testing and improving these resources, rather than choosing what we think is best based on observations of student behaviours alone.

(6) **Imposing or Appropriating Constraints to Make them Meaningful.** In the previous two examples, we have explained why relaxing constraints can be beneficial. We have also found that sometimes, it is useful to *impose meaningful constraints* when doing experiment-inspired design. For example, because our meta-skills interventions have to fit within students' packed and multi-faceted lives, it is useful for us to limit the duration of each meta-skills activity, and still try to make it useful. For instance, the stress-reappraisal intervention was a brief 5-minute activity embedded within several other activities and adapted from what was a much longer, 30-minute intervention done in a different context in prior literature. By forcing ourselves to shorten these interventions, we had to come up with more creative solutions to condense information and get to the point.

## 4 CONCLUSION

In this paper, we suggest that it would be valuable for researchers to explore a family of strategies that we may call *experiment-inspired design*. We share some ideas and approaches for members in this family in the form of *six lessons* that we draw from our experience with running several meta-skills interventions based on these strategies over the past several years. We believe this can be an an active area for researchers to explore more broadly in future, and deeply consider alternative ways to think about experiments, and in doing so, *experiment with experimentation.*

## 5 ACKNOWLEDGEMENTS

This work was partially supported by the Natural Sciences and Engineering Research Council of Canada (NSERC) (#RGPIN-2019-06968), as well as by the Office of Naval Research (ONR) (#N00014-18-1-2755 and #N00014-21-1-2576).